\documentclass[fleqn,10pt]{two-column} 
\usepackage[english]{babel} 
\usepackage{amsmath,amssymb}
\usepackage{xspace}
\usepackage{lettrine}
\usepackage{overpic}
\usepackage{upgreek}

\input{babarsym.tex} 

\newcommand{\Dx}{\ensuremath{D^{(*)}}\xspace}
\newcommand{\Btag}{\ensuremath{B_{\rm tag}}\xspace}
\newcommand{\Bsig}{\ensuremath{B_{\rm sig}}\xspace}

\def\mupm{\ensuremath{\mu^{\pm}}\xspace}

\def\Upsil{\ensuremath{\Upsilon(4S)}\xspace}

\def\BDtaunu{\ensuremath{\Bbar \rightarrow D \tau^-\nutb}\xspace}
\def\BDstaunu{\ensuremath{\Bbar\rightarrow D^* \tau^-\nutb}\xspace}

\def\BDellnu{\ensuremath{\Bbar \rightarrow D \ell^- \nulb}\xspace}
\def\BDsellnu{\ensuremath{\Bbar \rightarrow D^* \ell^- \nulb}\xspace}
\def\BDssellnu{\ensuremath{\Bbar \rightarrow D^{**} \ell^- \nulb}\xspace}

\def\BDxtaunu{\ensuremath{\Bbar \rightarrow D^{(*)} \tau^- \nutb}\xspace}
\def\BDxellnu{\ensuremath{\Bbar \rightarrow D^{(*)} \ell^- \nulb}\xspace}

\def\Bntaunu{\ensuremath{B^- \rightarrow \tau^- \nutb}\xspace}
\def\Btaunu{\ensuremath{B^- \rightarrow \tau^- \nutb}\xspace}
\def\Bellnu{\ensuremath{B^- \rightarrow \ell^- \nulb}\xspace}

\def\BRtaunu{\ensuremath{{\cal B}(B^- \rightarrow \tau^- \nutb)}\xspace}

\def\RDx{\ensuremath{{\cal R}_{\Dx}}\xspace}
\def\RDs{\ensuremath{{\cal R}_{D^*}}\xspace}
\def\RD{\ensuremath{{\cal R}_{D}}\xspace}
\def\Esl{\ensuremath{E^*_\ell}\xspace}

\def\mmiss{\ensuremath{m_{\rm miss}^2}\xspace}

\def\upmus{\ensuremath{\upmu \mathrm{s}}\xspace}

\definecolor{color1}{RGB}{0,0,90} 
\definecolor{color2}{RGB}{0,20,20} 

\usepackage{hyperref} 
\hypersetup{hidelinks,colorlinks,breaklinks=true,urlcolor=color2,citecolor=color1,linkcolor=color1,bookmarksopen=false,pdftitle={Title},pdfauthor={Author}}

\JournalInfo{Prepared for submission to Nature} 
\Archive{\today} 

\PaperTitle{A Challenge to Lepton Universality in B Meson Decays} 

\Authors{Gregory Ciezarek\textsuperscript{1}, Manuel Franco Sevilla\textsuperscript{2},
Brian Hamilton\textsuperscript{3}, Robert Kowalewski\textsuperscript{4}, 
Thomas Kuhr\textsuperscript{5}, Vera L\"uth\textsuperscript{6}, Yutaro Sato\textsuperscript{7}} 
\affiliation{\textsuperscript{1}\textit{Nikhef National Institute for Subatomic Physics, Amsterdam, The Netherlands}}
\affiliation{\textsuperscript{2}\textit{University of California at Santa Barbara, Santa Barbara, California 93106, USA}}
\affiliation{\textsuperscript{3}\textit{University of Maryland, College Park, Maryland 20742, USA}}
\affiliation{\textsuperscript{4}\textit{University of Victoria, Victoria, British Columbia, Canada V8P 5C2}}
\affiliation{\textsuperscript{5}\textit{Ludwig Maximilians University, 80539 Munich, Germany}}
\affiliation{\textsuperscript{6}\textit{SLAC National Accelerator Laboratory, Stanford, California 94309, USA}}
\affiliation{\textsuperscript{7}\textit{Kobayashi-Maskawa Institute, Nagoya University, Nagoya 464-8602, Japan}}

 \newcommand{\keywordname}{Keywords} 
\Abstract
{One of the key assumptions of the Standard Model of fundamental particles is that the interactions of the charged leptons, namely electrons, muons, and taus, differ {\em only} because of their different masses.
While precision tests comparing processes involving electrons and muons have not revealed any definite violation of this assumption, recent studies involving the higher-mass tau lepton have resulted in observations that challenge lepton universality at the level of four standard deviations. A confirmation of these results 
would point to new particles or interactions, and could have profound implications for our understanding of particle physics.}

\begin{document}
\flushbottom 
\maketitle 
\thispagestyle{empty} 


\label{sec:intro}

\lettrine[lines=3,findent=2pt]{\color{color1}M}{ }ore than
70 years of particle physics research have led to an elegant and concise theory of particle interactions at
the sub-nuclear level, commonly referred to as the Standard Model (SM)~\cite{Mann:2010zz,Weinberg:1996kr}.
Based on information extracted from experiments, theorists have combined the theory of electroweak (EW)
interactions with quantum chromodynamics (QCD), the theory of strong interactions, and experiments have
validated this theory to an extraordinary degree.  Any observation that is proven to be inconsistent with SM
assumptions would suggest a new type of interaction or particle.
 
In the framework of the SM of particle physics the fundamental building blocks, quarks and leptons, are each
grouped in three generations of two members each.  The three charged leptons, the electron ($e^-$), the muon
($\mu^-$) and the tau ($\tau^-$) are each paired with a very low mass, electrically neutral neutrino,
$\nue, \num,$ and $\nut$.  The electron, a critical component of matter, was discovered by J.J.
Thomson~\cite{Thomson} in 1897. The discovery of the muon in cosmic rays by C. D. Anderson and
S. H. Neddermeyer~\cite{Neddermeyer:1937md} in 1937 came as a surprise, similarly surprising was the first
observation of $\tautau$ pair production by M. Perl et al.~\cite{Perl:1975bf} at the SPEAR $\epem$ storage
ring in 1975. As far as we know, all leptons are point-like particles, i.e. they have no substructure.

The three generations are ordered by the mass of the charged lepton ranging from 0.511\mev for $\epm$ to
105\mev for $\mupm$, and 1,777\mev for $\taupm$~\cite{Ablikim:2014uzh}.  These mass differences lead to vastly
different lifetimes, from the stable electron to 2.2\upmus for muons, and 0.29\ps for taus.  Charged leptons
participate in electromagnetic and weak, but not strong interactions, whereas neutrinos only undergo weak
interaction.  The SM assumes that these interactions of the charged and neutral leptons are universal, i.e.,
the same for the three generations.

Precision tests of lepton universality have been performed over many years by many experiments.  To date no
definite violation of lepton universality has been observed.  Among the most precise tests is a comparison of
decay rates of $K$ mesons, $K^- \to e^-
\nueb$ versus $K^-\to \mu^- \numb$~\cite{Lazzeroni:2012cx}~\cite{charge}.
Furthermore, taking into account precision measurements of the tau and muon masses and lifetimes and the decay
rates $\taum \to e^- \nueb \nut$ and $\mu^- \to e^- \nueb \num$, the equality of the weak coupling strengths
of the tau and muon was confirmed~\cite{Ablikim:2014uzh}.  On the other hand, a recent determination of the
proton radius, derived from very precise measurements of the Lamb shift in muonic hydrogen
atoms~\cite{Pohl:2010zza} differs by about 4\% from measurements of normal hydrogen atoms and e-p scattering
data.  Studies of the origin of this puzzling difference are underway~\cite{Pohl:2013yb}. They are aimed at a
better understanding of the proton radius and structure, and may reveal details of the true impact of muons
and electrons on these interactions.

Recent studies of purely leptonic and semileptonic decays of $B$ mesons of the form $\Bntaunu$ and
$\BDxellnu$, with $\ell = e, \mu,$ or $\tau$, have resulted in observations that seem to challenge lepton
universality.  These weak decays involving leptons are well understood in the framework of the SM, and
therefore offer a unique opportunity to search for unknown phenomena and processes involving new particles,
for instance, a yet undiscovered charged partner of the Higgs boson~\cite{Tanaka:1994ay}.  Such searches have
been performed on data collected by three different experiments, the LHCb experiment at the proton-proton
($pp$) collider at CERN in Europe, and the BABAR and Belle experiments at $\epem$ colliders in the U.S.A. and
in Japan.

Measurements by these three experiments favor larger than expected rates for semileptonic $B$ decays involving
$\tau$ leptons.  Currently, the combined significance of these results is at the level of four standard
deviations, and the fact that all three experiments report an unexpected enhancement has drawn considerable
attention.  A confirmation of this violation of lepton universality and an explanation in terms of new physics
processes are a very exciting prospect! In the following, details of the experimental techniques and
preliminary studies to understand the observed effects will be presented, along with prospects for improved
sensitivity and complementary measurements at current and future facilities.



\subsection*{Standard model predictions of B meson decay rates}
\label{sec:bdecays}

\begin{figure}[btp!]
\centering
\includegraphics[width=0.34\textwidth]{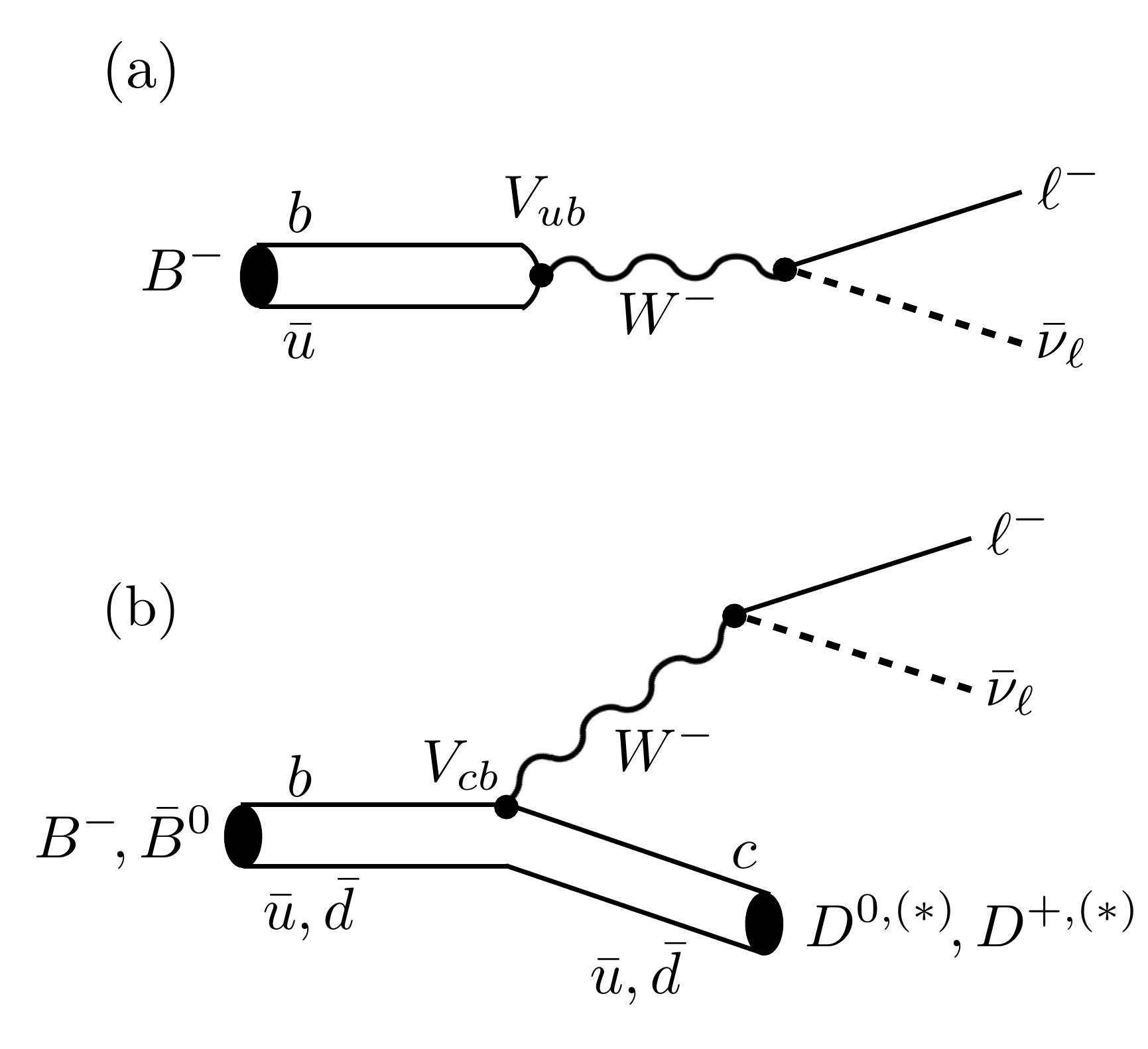}
\caption{{\bf Diagrams for SM decay processes:} (a) \Bellnu with a purely leptonic final state and (b) \BDxellnu)
involving a charm meson and lepton pair and mediated by a vector boson $(W^-$). }
\label{fig:feynman}
\end{figure}

According to the SM, purely leptonic and semileptonic decays of $B$ mesons are mediated by the $W^-$ boson, as shown schematically in
Figure~\ref{fig:feynman}.  $B$ mesons are assumed to be composed of a b-quark and an anti-quark, either
$\Bub(b,\ubar)$ or $\Bzb(b,\dbar)$, whereas charm mesons (the spin-0 $D$ and spin-1 $D^*$ state) contain a
c-quark and an anti-quark, $D^{0(*)}(c,\ubar)$ or $D^{+(*)}(c,\dbar)$.

For purely leptonic $\Bbar$ decays, the SM prediction of the total decay rate $\Gamma$, which depends critically 
on the lepton mass squared $ m^2_{\ell}$, is
\begin{equation}
\Gamma^{SM}(\Bellnu) =  
\frac{G_F^2\; m_B \; m^2_{\ell}}{8\pi}|V_{ub}|^2
  \left(1-\frac{m_\ell^2}{m_B^2} \right)^2 \times f^2_B .  
\label{eq:Gamma_pl}
\end{equation}
\noindent
The first factor contains the Fermi constant $G_F=1.1663797 \times 10^{-5} \gev^{-2}$ and the $\B$ meson mass, $m_B=5.279\gev$.
All hadronic effects, due to the binding of quarks inside the meson, are encapsulated in the decay constant $f_B$. Recent lattice QCD calculations~\cite{Aoki:2013ldr} predict $f_B=(0.191 \pm
0.009)\gev$.  Taking into account the current world averages for the $B^-$ lifetime, $\tau_B=(1.638\pm
0.004)$\ps~\cite{Agashe:2014kda}, and the quark mixing parameter~\cite{Kobayashi:1973fv} for $b \to u$ transitions$, |V_{ub}|$~\cite{Amhis:2014hma}, 
 the expected branching fraction, i.e., the frequency of this decay relative to all decay modes, is~\cite{Charles:2004jd}
\begin{equation}
{\cal B}^{SM} (\Bm \to \taum \nutb)=(0.75~^{+0.10}_{-0.05}) \times 10^{-4}.
\end{equation}
\noindent
Decays to the lower mass charged leptons, $e^-$ and $\mu^-$, are strongly suppressed by spin effects and have not yet been observed.

The differential decay rate, ${\rm d}\Gamma$, for semileptonic decays involving $D^{(*)}$ mesons depends on both $\m^2_{\ell}$ and $q^2$, the invariant mass squared of the lepton 
pair~\cite{Korner:1989qb},  
%
\begin{align}
&\frac{{\rm d}\Gamma^{SM}(\BDxellnu)}{{\rm d}q^2}\,   
=  \underbrace{\frac{G_F^2\; |V_{cb}|^2\; |\boldsymbol{p}^*_{D^{(*)}}| \; q^2}{96\pi^3 m_B^2}
  \left(1-\frac{m_\ell^2}{q^2} \right)^2}_{\text{universal and phase space  factors}}
 \\ \nonumber & \times 
\underbrace{\left[(|H_{+}|^2+|H_{-}|^2+|H_{0}|^2) \left(1+\frac{m_\ell^2}{2q^2}  \right) + \frac{3 m_\ell^2}{2q^2}|H_{s}|^2 \right]}_{\text{hadronic effects}}~.
 \label{eq:Gamma_sl}
\end{align}
\noindent
The first factor is universal for all semileptonic $B$ decays, containing a quark flavor mixing parameter~\cite{Kobayashi:1973fv}, in this case $|V_{cb}|$~\cite{Amhis:2014hma} for $b \to c$ quark transitions, and $p^*_{D^{(*)}}$,
the 3-momentum of the hadron in the $B$ rest frame, in this case a $D^{(*)}$ meson. The four helicity~\cite{helicity} amplitudes $H_+, H_-, H_0$ and $H_s$ capture the impact of hadronic
effects. They depend on the spin of the charm meson and on $q^2$.  The kinematic range, $m^2_{\ell} \le q^2 \le (m_B - m_{D^{*}})^2$, is sensitive to the lepton mass $m_{\ell}$ and the charm meson mass $m_{D^{*}}$.
The much larger mass of the $\tau$ not only impacts the rate, but also the kinematics of the decays
via the $H_s$ amplitude. All four amplitudes contribute to $\BDsellnu$, 
while only  $H_0$ and $H_s$ contribute to $\BDellnu$, which leads to a higher sensitivity of this decay mode to the scalar contribution $H_s$.

Measurements of the ratios of semileptonic branching fractions remove the dependence on $|V_{cb}|$, lead to a partial cancellation of theoretical uncertainties related to hadronic effects, and reduce of the impact of experimental uncertainties. Current SM predictions~\cite{Na:2015kha, Fajfer:2012vx, Lattice:2015rga} are
\begin{eqnarray}
\label{eq:RD}
{\cal R}^{SM}_D     &=& \frac {{\cal B}(\Bbar \to D \tau^- \nutb)}
                              {{\cal B}(\Bbar \to D e^- \nueb)} 
= 0.300 \pm 0.008 \\
\label{eq:RDs}
{\cal R}^{SM}_{D^*} &=&\frac {{\cal B}(\Bbar \to D^* \tau^- \nutb)}
                             {{\cal B}(\Bbar \to D^* e^- \nueb)} 
= 0.252 \pm 0.003 . \
\end{eqnarray}
\noindent
The predicted ratios relative to ${\cal B}(\Bbar \to D^* \mu^- \numb)$ are identical within the quoted precision.

\subsection*{B meson production and detection}
\label{sec:experiments}

$\B$ meson decays have been studied at $pp$ and $e^+e^-$ colliding beam facilities, operating at very
different beam energies. The $e^+e^-$ colliders operated at a fixed energy of 10.579\gev in the years 1999 to
2010.  At this energy, about 20 \mev above the kinematic threshold for $\BB$ production, $e^+$ and $e^-$
annihilate and produce a particle,
commonly refered to as \Upsil, which decays almost exclusively to $\BpBm$ or $\BzBzb$ pairs.  The maximum
production rate for these $\Upsil\to \BB$ events of 20~Hz was achieved at KEK, compared to the multi-hadron
non-$\BB$ background rate of about 80~Hz.
 
\begin{figure*}[t!]
\centering
\includegraphics[width=0.9\textwidth]{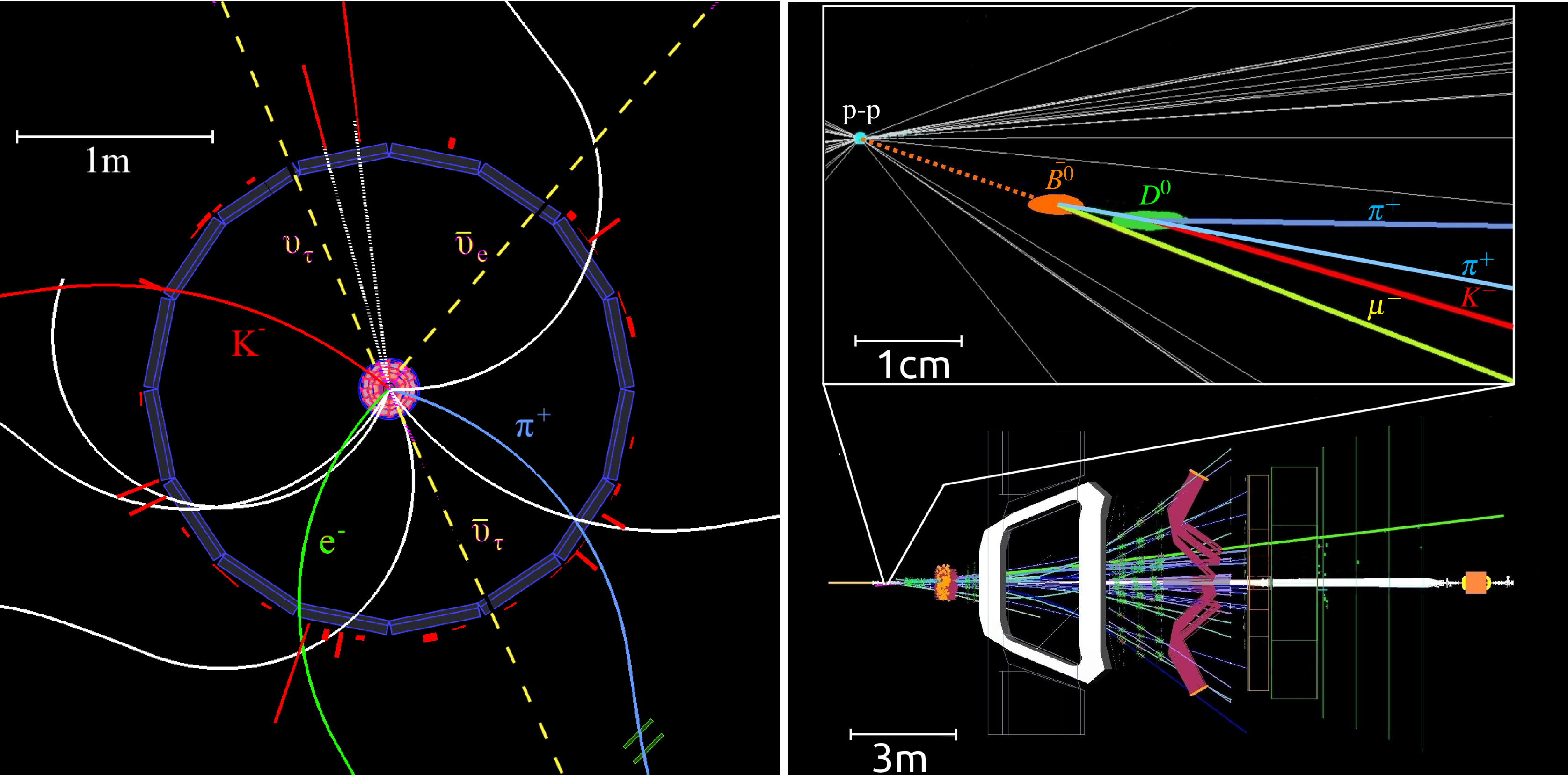}
\vspace{3mm}
\caption{ {\bf Belle (a) and LHCb (b) single event displays} illustrating the reconstruction of semileptonic 
$\B$ meson decays: Trajectories of charged particles are shown as colored solid lines, energy deposits in the
calorimeters are depicted by red bars.  The Belle display is an end view perpendicular to the beam axis with
the silicon detector in the center (small orange circle) and the device measuring the particle velocity (dark
purple polygon).  This is a $\Upsil \to B^+ B^-$ event, with $B^- \to D^0 \tau^- \bar{\nu_\tau}$, $D^0 \to
K^- \pi^+$ and $\tau^- \to e^- \nu_\tau \bar{\nu_e}$, and the $B^+$ decaying to five charged particles (white
solid lines) and two photons.  The trajectories of undetected neutrinos are marked as dashed yellow lines.
The LHCb display is a side view with the proton beams indicated as a white horizontal line with the
interaction point far to the left, followed by the dipole magnet (white trapezoid) and the Cherenkov detector
(red lines). The area close to the interaction point is enlarged above, showing the tracks of the charged
particles produced in the $pp$ interaction, the $\Bz$ path (dotted orange line), and its decay
$\bar{B^{0}} \to D^{*+} \tau^{-} \bar{\nu}_\tau$ with $D^{*+} \to D^0\pi^+$ and $D^0 \to K^- \pi^+$, plus the
$\mu^-$ from the decay of a very short-lived $\tau^{-}$.  }
\label{fig:events_displays}
\end{figure*}

$\B$ mesons from \Upsil decays have very low momenta, $\approx 300$\mev, and therefore their decay products
are distributed almost isotropically in the rest frame of the \Upsil. For this reason, the
BABAR~\cite{Aubert:2001tu,TheBABAR:2013jta} and Belle~\cite{Abashian:2000cg} detectors were designed to cover
close to 90\% of the total solid angle, thereby enabling the reconstruction of all final state particles from
decays of the two $B$ mesons, except for neutrinos. Both detectors consist of cylindrical layers of sensors
surrounding the beam pipe, plus endcaps to cover the small polar angles.  Constraints from energy-momentum
conservation allow events containing a single neutrino to be discriminated from those containing multiple
undetected particles. This feature also allows for an effective suppression of non-$\BB$ background and
misreconstructed events.

The LHC $pp$ collider operated at total energies of 7 and 8 TeV from 2008 to 2012.  In inelastic $pp$
collisions, high energy gluons, the carriers of the strong force between the quarks inside the protons,
collide and produce a pair of $B$ hadrons (mesons or baryons) along with a large number of other charged and
neutral particles, in roughly one of hundred $pp$ interactions.  The $B$ hadrons are typically produced at
small angles to the beam and with high momenta, features that determined the design of the LHCb
detector~\cite{Alves:2008zz,Dettori:2013xsa}, a single arm forward spectrometer, covering the polar angle
range of $3-23$ degrees.  The high momentum and relatively long $B$ hadron lifetime result in decay distances
of several cm. Very precise measurements of the $pp$ interaction point, combined with the detection of charged
particle trajectories from $\B$ decays which do not intersect this point, are the very effective, primary
method to separate $\B$ decays from background.

All three experiments rely on several layers of finely segmented silicon strip detectors to locate the
beam-beam interaction point and decay vertices of long-lived particles.  A combination of silicon strip
detectors and multiple layers of gaseous detectors measure the trajectories of charged particles, and
determine their momenta from the deflection in a magnetic field. Examples of reconstructed signal events
recorded by the LHCb and Belle experiments are shown in Figure~\ref{fig:events_displays}.

For a given momentum, charged particles of different masses, primarily pions and kaons, are identified by
their different velocities.  All three experiments make use of devices which sense Cherenkov radiation,
emitted by particles with velocities that exceed the speed of light in a chosen radiator material. For lower
velocity particles, Belle complements this with time-of-flight measurements.  BABAR and Belle also measure the
velocity-dependent energy loss due to ionization in the tracking detectors.  Arrays of cesium iodide crystals
measure the energy of photons and identify electrons in BABAR and Belle.  Muons are identified as particles
penetrating a stack of steel absorbers interleaved with large area gaseous detectors.

\subsection*{Measurements of \texorpdfstring{\Btaunu}{B -> TauNu} decays}
\label{sec:taunu}

The decays
$\Bm \to \taum \nutb$ with two or three neutrinos in the final state have only been observed by BABAR and Belle.   
These two experiments exploit the $\BB$ pair production at the $\Upsil$ resonance via the process 
$e^+ e^- \to \Upsil \to \BB$. These $\BB$ pairs can be tagged by the 
reconstruction of a hadronic or semileptonic decay of one of the two $\B$ mesons, referred to as $B_\mathrm{tag}$.
If this decay is correctly reconstructed, all remaining particles in the event originate from the other $\B$ decay.

BABAR and Belle have independently developed two sets of algorithms to tag 
$\BB$ events. 
The hadronic tag algorithms~\cite{Feindt:2011mr,Lees:2013uzd} search for the best match between one of more than a thousand possible decay chains and a subset of all detected particles in the event. 
The efficiency for finding a correctly matched $B_\mathrm{tag}$ is unfortunately quite small, 0.3\%.
The benefit of reconstructing all final state particles is that the total energy, $E_\mathrm{miss}$, and vector momentum, $\vec{p}_\mathrm{miss}$,
of all undetected particles of the other $B$ decay can be inferred from energy and momentum conservation.
The invariant mass squared of all undetected particles, $m_\mathrm{miss}^2 =  E_\mathrm{miss}^2 - \vec{p}_\mathrm{miss}^2$,
is used to distinguish events with one neutrino ($m_\mathrm{miss}^2 \approx 0$) from events with multiple neutrinos or other missing particles ($m_\mathrm{miss}^2 > 0$).

The semileptonic tag algorithms exploit the large branching fractions for 
$\B$ decays involving a charm meson, a charged lepton and associated neutrino,
$\B \to D^{(*)} \ell^+ \nul$, with $\ell^+=e^+, \mu^+$. 
The efficiency for finding these tag decays is about 1\%.
However, the presence of the neutrino leads to weaker constraints on the $B_\mathrm{tag}$ and signal $\B$  decay.

Measurements of $\Bm \to \taum \nutb$ decays are based on leptonic $\tau$ decays, $\taum\to e^-\nueb\nut$ and $\taum\to \mu^-\numb\nut$, and on semileptonic decays, $\taum\to \pi^- \nut$ and $\taum\to \pi^- \pi^0\nut$, which together account for 70\% of all 
$\tau^-$ decays. 
Thus, the signature for signal events is a single charged particle, either a charged lepton, a $\pi^-$, or a $\pi^-$ accompanied by a $\pi^0$, plus a $B_\mathrm{tag}$. 
 
\begin{figure}[btp!]
\centering
\includegraphics[width=0.4\textwidth]{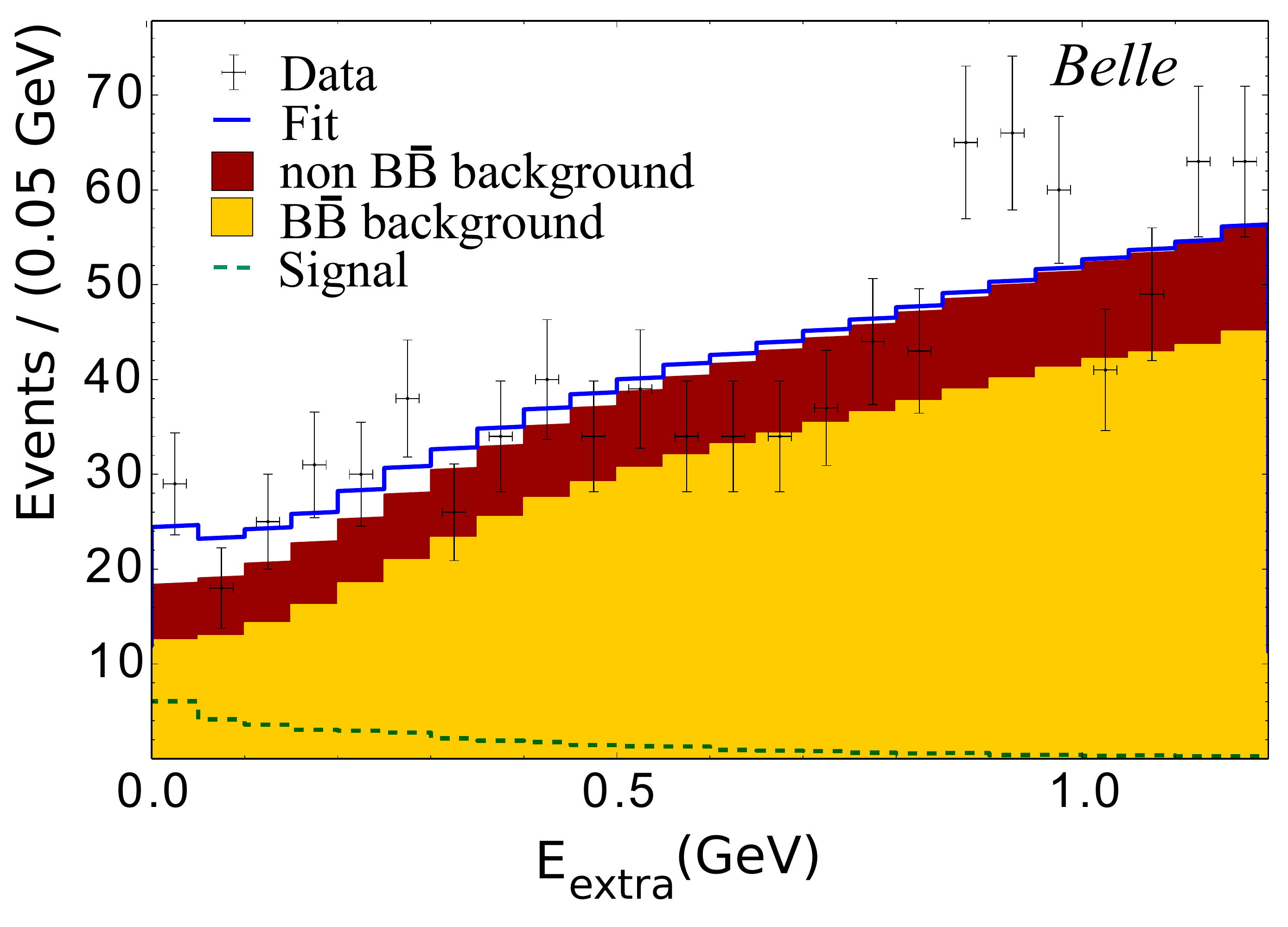}
\caption{{\bf Extraction of the $\Bm \to \taum \nutb$  yield from Belle data:} 
Results of a fit to the $E_\mathrm{extra}$ distribution for the sum of  
signal and backgrounds~\cite{Kronenbitter:2015kls} for a subset of events with 
$\taum\to \pi^- \nut$ candidates.
The green histogram at the bottom indicates the predicted signal distribution.}
\label{fig:Eextra}
\end{figure}

The presence of multiple neutrinos precludes the use of kinematic constraints to effectively suppress backgrounds from other $\B$ decays.
A variable that is sensitive to backgrounds with additional photons or undetected charged particles due to efficiency and acceptance losses  
is $E_\mathrm{extra}$, the sum of the energy deposits in the calorimeter which are not associated with the tag or signal $B$ decay.
Figure~\ref{fig:Eextra} shows a $E_\mathrm{extra}$
distribution measured by Belle for a subset of events with $\taum\to \pi^-\nut$. Signal events have low values of $E_\mathrm{extra}$,
while background events extend to higher values.
The signal yield is determined from a fit to the data using signal and background distributions based on data control
samples and Monte Carlo simulation.  
The sum of the fitted signal yields for the four subsamples of purely leptonic and semileptonic $\tau$ decays, corrected for the efficiency of the tag and signal $\B$  decays, is used to determine the $\Bm \to \taum \nutb$ branching fraction. 

\begin{figure*}[btp!]
\centering
\includegraphics[width=0.9\textwidth]{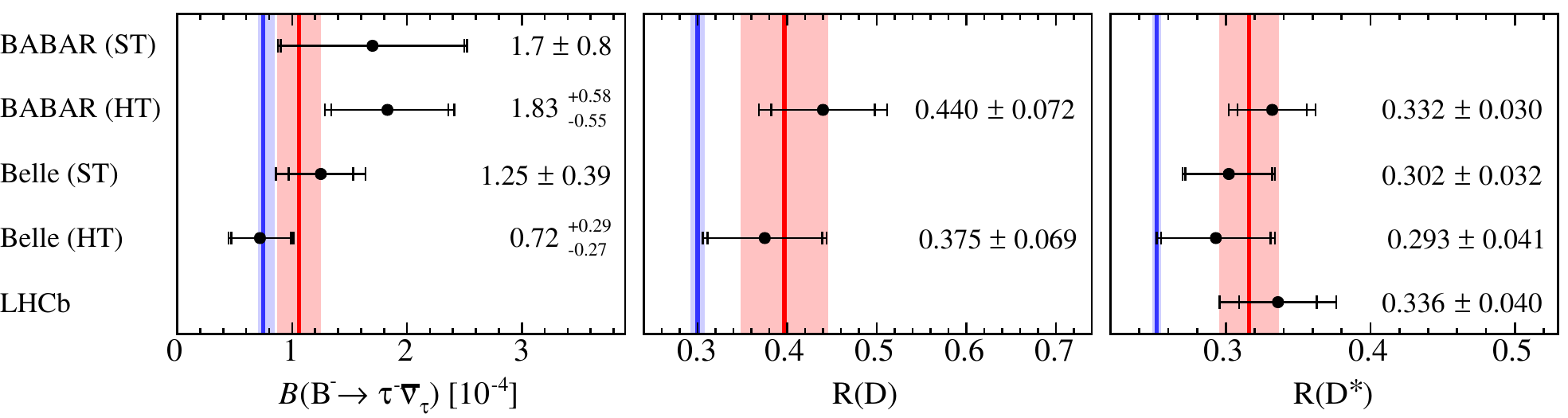}
\caption{{\bf Comparison of measurements with SM predictions:} The branching fraction \BRtaunu (left), the ratios \RD (center), and \RDs (right) by 
BABAR~\cite{Aubert:2009wt, Lees:2012ju,Lees:2013uzd}, 
Belle~\cite{Kronenbitter:2015kls, Adachi:2012mm, Huschle:2015rga,Sato:2016svk},
and LHCb~\cite{Aaij:2015yra}.
The data points indicate statistical and total uncertainties.
ST and HT refer to the measurements with semileptonic and hadronic tags, respectively.
The average values of the measurements and their combined uncertainties, obtained by the Heavy Flavor Averaging 
Group~\cite{HFAG2016, HFAG2016:RDx}, are shown in red as vertical lines and bands, and the expectations from the SM calculations~\cite{Charles:2004jd,Na:2015kha,Fajfer:2012vx} are shown in blue.}
\label{fig:results}
\end{figure*}

As shown in Figure~\ref{fig:results}, current measurements by Belle~\cite{Kronenbitter:2015kls, Adachi:2012mm} and BABAR~\cite{Aubert:2009wt, Lees:2012ju} are of limited
precision due to very small signal samples and high backgrounds,
and uncertainties in the $B_\mathrm{tag}$ efficiencies.
The combination of these four measurements constitutes the first 
observation of a purely leptonic \Bm decay.
While the early measurements favored somewhat larger values,
the current average~\cite{HFAG2016} of
\begin{equation}
{\cal B} (\Bm \to \taum \nutb)=(1.06 \pm 0.19) \times 10^{-4}.
\end{equation}
\noindent
is compatible with the SM prediction~(compare Eq. 1), which is lower
by $1.4$ standard deviations.

\subsection*{Measurements of \texorpdfstring{\BDxtaunu}{B -> D(*)TauNu} decays}
\label{sec:dxtaunu}
As defined in Eqs.~\ref{eq:RD} and \ref{eq:RDs}, \RDx correspond to the ratio of branching fractions for
\BDxtaunu (signal) and \BDxellnu (normalization).
BABAR and Belle events containing such decays are selected by requiring a hadronic \Btag, a $D$ or $D^*$
meson, and a charged lepton $\ell^-=e^-, \mu^-$.
Charged and neutral $D$ candidates are reconstructed from combinations of pions and
kaons with invariant masses compatible with the $D$ meson mass.  The higher-mass 
$D^{*0}$ and $D^{*+}$ mesons
are identified by their $D^{*} \rightarrow D\pi$ and $D^{*} \rightarrow D\gamma$ decays.  In signal decays,
the lepton $\ell^-$ originates from the $\tau^- \to \ell^- \nut \nulb$ decay, leading to a final state with three neutrinos and
resulting in a broad \mmiss distribution, while in normalization decays the lepton originates from the
$B$ decay with a single neutrino and therefore $\mmiss \approx 0$. Non-$B\bar{B}$ backgrounds and misreconstructed events are greatly suppressed by the \Btag reconstruction.
The remaining background is further reduced by multivariate selections.

At LHCb, only decays of $\bar{B}^0$ mesons producing a $\mu^-$ and $D^{*+}$ meson are selected. Muons are favored over electrons because of their higher detection efficiency and momentum resolution. The
$D^{*+}$ meson is reconstructed exclusively in $D^{*+} \to D^0 (\to K^- \pi^+) \pi^+ $ decays.
$B$ mesons produced at LHCb have a flight path of order 1 cm. This feature is exploited to reject the bulk of the background, by requiring that the charged particles from the $B$ candidate and no other tracks originate from a common vertex that is significantly separated from the $pp$ collision point.
The reduction in signal efficiency due to the use of a single decay chain is compensated by the very large production rate of $B$ mesons at the LHC.  
The direction of the $\B$ momentum is inferred from the reconstructed $pp$
collision point and $D^{*+}\mu^-$ vertex, its magnitude is unknown. LHCb approximates the $\B$ momentum
by equating its component parallel to the beam axis to that of the $D^{*+}\mu^-$ combination, rescaled by the
ratio of the $\B$ mass to the measured $D^{*+}\mu^-$ mass.

\begin{figure*}[bt!]
\centering
\includegraphics[width=0.95\textwidth]{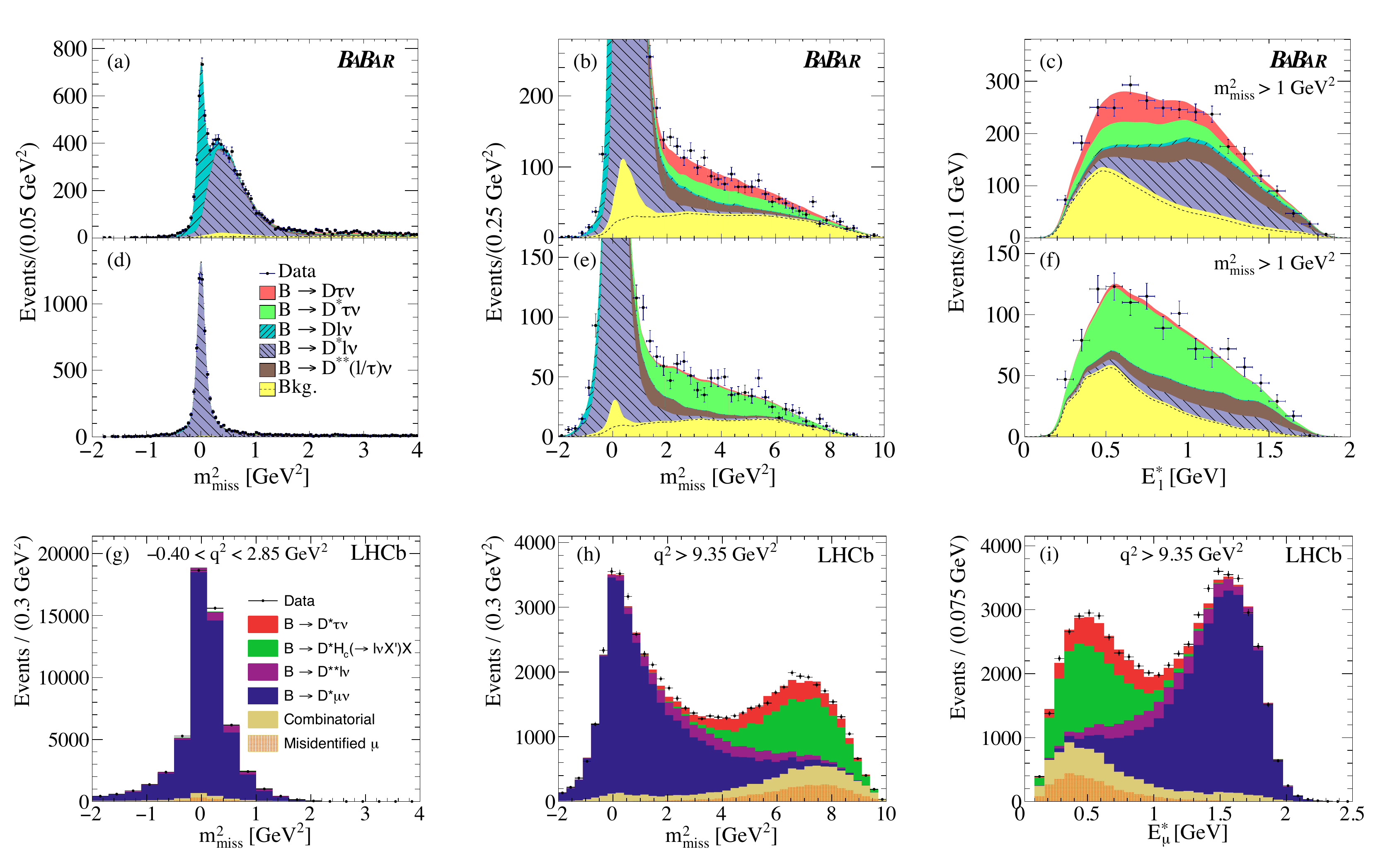}
\caption{ {\bf Extraction of the ratios ${\cal R}_D$ and ${\cal R}_{D^*}$ 
by maximum likelihood fits:}
Comparison of the projections of the measured \mmiss and \Esl distibutions (data points) 
and the fitted distributions of signal and background contributions for the BABAR 
fit~\cite{Lees:2013uzd} to the $D\ell$ samples (a-c) and $D^*\ell$ samples (d-f), as well the
LHCb fit~\cite{Aaij:2015yra} to the $D^{*+}\ell$ sample (g-i).
The $D\ell$ samples in (a-c) show sizable contributions from $B^0 \to D^{*+}\ell^- \nulb$ and  $B^0 \to D^{*+}\tau^- \nutb$ decays, where the low energy pion or photon originating from a $D^* \to D \pi$ or $D^* \to D \gamma$ transition was undetected.  
The BABAR data exclude  $q^2 <4\text{ GeV}^2$, where the contributions from signal decays is very small.
The \Esl distributions in  (c) and (f) are signal enhanced by the restriction $\mmiss > 1 \gev^2$.  
The  LHCb results are presented for two different $q^2$ intervals, the lowest, which is free of 
$B^0 \to D^{*+}\tau^- \nutb$ decays (g), and the highest where this contribution is large (h,i).}
\label{fig:fit-dxtaunu}
\end{figure*}

The yields for the signal, normalization, and various background contributions are determined by maximum likelihood
fits to the observed data distributions. 
Control samples are used to
validate the simulated distributions and constrain the size and kinematic features of the background contributions.

All three experiments rely on the variables \mmiss, \Esl, the energy of the charged lepton
in the $B$ rest frame, and $q^2$.
BABAR and Belle restrict the data to $q^2>4\text{ GeV}^2$ to enhance the contribution from signal decays.
BABAR performs the fit in two dimensions whereas LHCb covers the whole $q^2$
range in four intervals, thus performing a fully three-dimensional fit.  
Belle performs a one-dimensional fit to the \mmiss distribution in the low \mmiss region ($\mmiss < 0.85\text{ GeV}^2$) dominated by the normalization decays, combined with a fit to
a multivariate classifier in the high \mmiss region.
This classifier includes \mmiss, \Esl, $E_{\rm extra}$, and additional kinematic variables.

Figure~\ref{fig:fit-dxtaunu} shows one-dimensional projections of the data and the
fitted contributions from signal, normalization, and backgrounds decays.  For BABAR (and likewise for Belle)
the \mmiss distributions show a narrow peak at zero (Figure~\ref{fig:fit-dxtaunu} a,d), dominated by
normalization decays with a single neutrino, whereas the signal events with three neutrinos extend to about 10 GeV$^2$. For \BDellnu decays, there is a sizable contribution from \BDsellnu decays, for which the pion or
photon from the $D^*\to D\pi$ or $D^*\to D\gamma$ decay was not reconstructed.  For LHCb, the peak at zero is
somewhat broader and has a long tail into the signal region (Figure~\ref{fig:fit-dxtaunu} h) because of
the sizable uncertainty in the estimation of the \Bsig momentum.  
The \Esl distributions (Figure~\ref{fig:fit-dxtaunu} c,f,i) provide
additional discrimination, since a lepton from a normalization decay has a higher average momentum than a lepton
originating from secondary $\tau^- \to \ell^- \nut \nulb$ decay in a signal $B$ decay.

Among the background contributions, semileptonic $B$ decays to $D^{**}$ mesons (charm mesons of higher mass than the $D^{*}$ mesons) are of concern, primarily because their branching fractions are not well known. These $D^{**}$ states decay to a $D$ or $D^{*}$ meson plus additional particles that, if not reconstructed,
contribute to the missing momentum of the decay. As a result, \BDssellnu decays have a broader $\mmiss$
distribution than normalization decays. They can be distinguished from signal decays by their \Esl distributions which extend to higher values.  
At LHCb, an important background arises from $B\to D^{(*)}H_c X$ decays, where $H_c$ is a charm hadron decaying either leptonically or semileptonically, and $X$ refers to additional low mass hadrons, if present. These decays produce \mmiss and \Esl spectra that are similar to those of signal events (Figure~\ref{fig:fit-dxtaunu} h,i). 

\begin{figure}[btp!]
\centering
\includegraphics[width=0.48\textwidth]{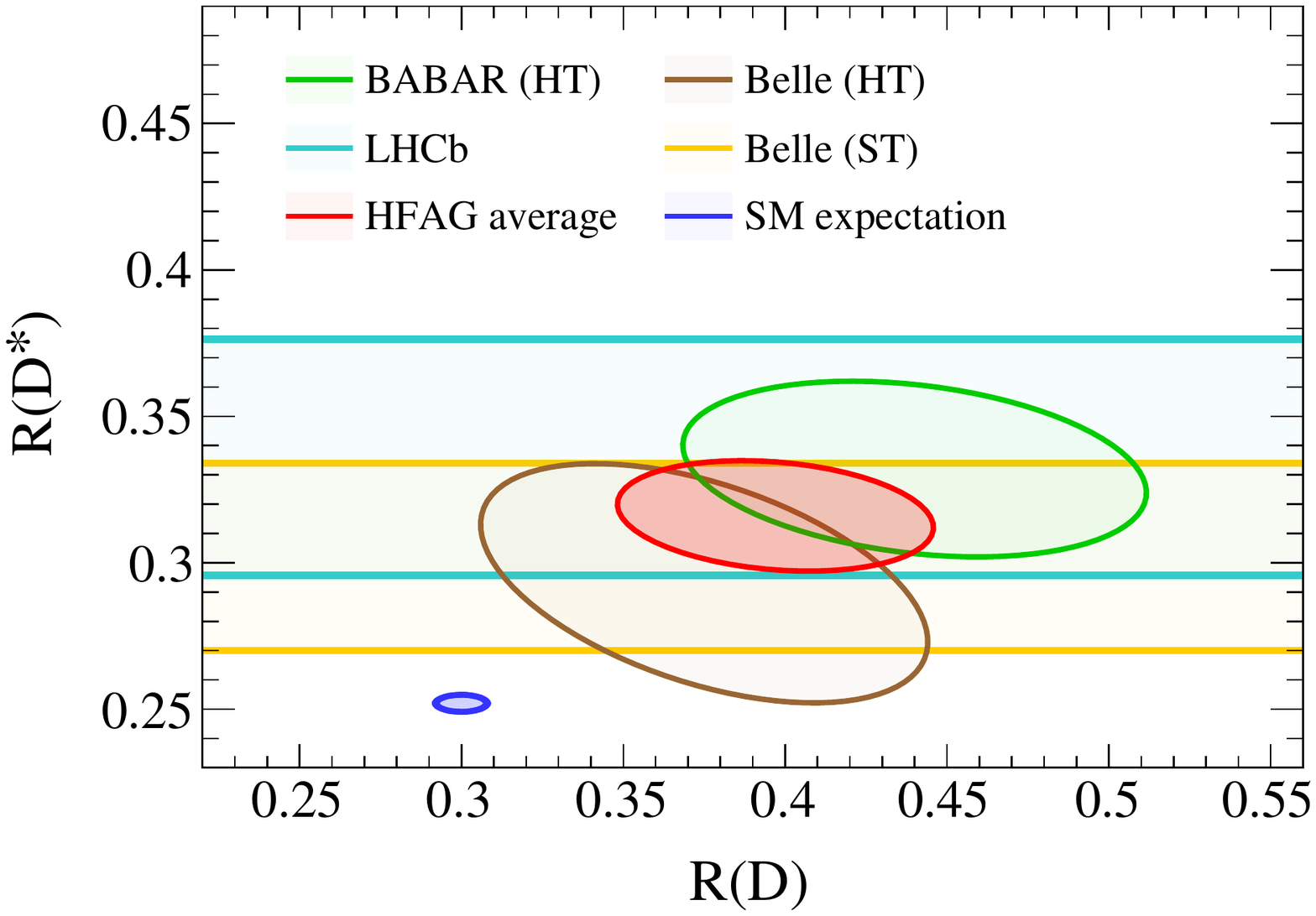}
\caption{ {\bf \RD and \RDs measurements:}  Results from
BABAR~\cite{Lees:2013uzd}, Belle~\cite{Huschle:2015rga, Sato:2016svk}, 
and LHCb~\cite{Aaij:2015yra}, their values and 1-$\sigma$ contours. The
average calculated by the Heavy Flavor Averaging Group~\cite{HFAG2016:RDx} is compared to SM predictions~\cite{Na:2015kha,Fajfer:2012vx,Lattice:2015rga}. 
ST and HT refer to the measurements with semileptonic and hadronic tags, respectively.}
\label{fig:results-dxtaunu-hfag}
\end{figure}

Figure~\ref{fig:results} shows the measured values for \RD and \RDs by BABAR~\cite{Lees:2013uzd},
Belle~\cite{Huschle:2015rga,Sato:2016svk}, 
and LHCb~\cite{Aaij:2015yra}. These results include 
a recent measurement of \RDs by Belle that employs a semileptonic tag, but do not include earlier results from BABAR~\cite{Aubert:2007dsa,Aubert:2009at} 
and Belle~\cite{Matyja:2007kt, Bozek:2010xy} based on partial data sets.
The averages of the measurements~\cite{HFAG2016:RDx} are 
\begin{eqnarray}
{\cal R}_D     &=& 0.397 \pm 0.040_{\rm stat} \pm 0.028_{\rm syst}, \\
{\cal R}_{D^*} &=& 0.316 \pm 0.016_{\rm stat} \pm 0.010_{\rm syst}.
\end{eqnarray}
Both values exceed the SM expectations. Taking into account the correlations (Figure~\ref{fig:results-dxtaunu-hfag}), 
the combined difference between
the measured and expected values has a significance of about four standard deviations.

\subsection*{Interpretations of results}
\label{sec:interpretations}

The results presented here have attracted the attention of the physics community, and have resulted in several
potential explanations of this apparent violation of lepton universality for decays involving the $\tau$
lepton.

In the SM, these $\B$ decays are mediated by a virtual charged vector boson, a particle of spin 1, usually
referred to as the $W^-$ (as indicated in the diagram in Figure 1) which couples equally to all leptons.  
If a hitherto unknown virtual particle existed that interacted differently with leptons of higher
mass like the $\tau^-$, this could change the $\B$ decay rates and their kinematics.

Among the simplest explanations for the observed rate increases for decays involving
$\tau^-$ would be
the existence of a new vector boson, $W'^-$, similar to the SM $W^-$ boson, but with a greater mass,
and with couplings of varying strengths to different leptons and quarks.  This could lead to changes
in \RD\ and \RDs, but not in the kinematics of the decays, which are observed to be consistent with the SM.
However, this choice is constrained by searches for $W'^- \to t\bar{b}$
decays~\cite{Chatrchyan:2012gqa,Aad:2014xea} at the LHC collider at CERN, as well as by precision measurements
of $\mu$~\cite{Prieels:2014paa} and $\tau$~\cite{Stahl:2000aq} decays.

Another potentially interesting candidate would be a new type of Higgs boson, a particle of spin 0, similar to the recently discovered neutral Higgs~\cite{Chatrchyan:2012ufa,Aad:2012tfa}, but
electrically charged. 
This charged Higgs ($H^-$) was proposed in
minimal extensions of the SM~\cite{Barger:1989fj}, 
which are part of 
broader theoretical frameworks such as supersymmetry~\cite{Gunion:1984yn}.  The $H^-$ would mediate weak
decays, similar to the $W^-$ (as indicated in Figure 1), but couple differently to leptons of different mass. The $q^2$ and angular distributions would be impacted by this kind of mediator because of its different spin.

\begin{figure}[btp!]
\centering
\includegraphics[width=0.34\textwidth]{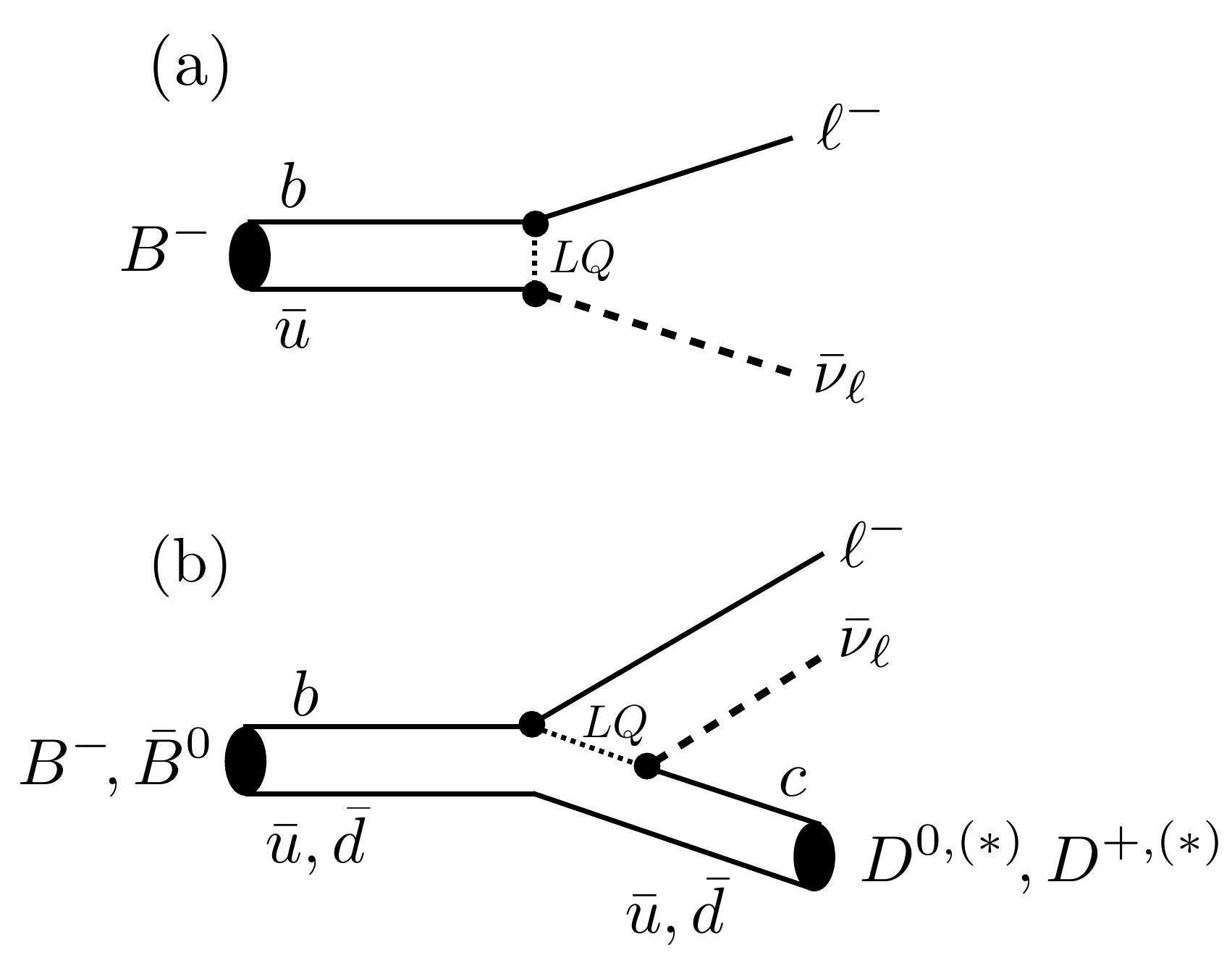}
\caption{ {\bf Diagrams for non-SM decay processes:} (a) \Bellnu with a purely leptonic final state and (b) \BDxellnu , 
involving a charm meson and lepton pair and mediated by a spin-0 lepto-quark~(LQ). }
\label{fig:feynman_lq}
\end{figure}

Another feasible solution might be leptoquarks~\cite{Dorsner:2016wpm}, hypothetical particles with both electric and color (strong) charges that allow transitions from 
quarks to leptons and vice versa, and offer a unified description of three generations of quarks and leptons.
Among the ten different types of leptoquarks, six could contribute to
$B \rightarrow D^{(*)} \tau \nu$ decays~\cite{Freytsis:2015qca}. A diagram of a spin-0 state mediating quark-lepton
transitions is shown in Figure~\ref{fig:feynman_lq} for the $\B$ decay modes under study.

BABAR and Belle have studied the implications of these hypothetical particles in the context of specific
models~\cite{Lees:2013uzd,Huschle:2015rga}.  The measured values of \RD\ and \RDs\ do not support the simplest
of the two-Higgs doublet models (type II), however, more general Higgs models with appropriate parameter
choices can accommodate these values~\cite{Datta:2012qk,Crivellin:2012ye,Fajfer:2012jt}.
Some of the leptoquark models could also explain the
measured values of \RD\ and \RDs~\cite{Sakaki:2013bfa,Dumont:2016xpj,Bauer:2015knc}, evading constraints from direct searches
of leptoquarks in $ep$ collisions~\cite{Buchmuller:1986zs} at HERA~\cite{Chekanov:2003af,Collaboration:2011qaa}
and $pp$ collisions at LHC~\cite{ATLAS:2013oea,Khachatryan:2014ura}. 

The three-body kinematics of $B \rightarrow D^{(*)} \tau \nu_{\tau}$ decays should permit further
discrimination of new physics scenarios based on the decay distributions of final state particles.  The
$q^2$ spectrum~\cite{Lees:2013uzd,Huschle:2015rga} and the momentum distributions of the $D^{(*)}$
and electron or muon~\cite{Sato:2016svk} have been examined.  Within the uncertainties of existing
measurements, the observed shapes of these distributions are consistent with SM predictions.

\subsection*{Conclusions and outlook}
\label{sec:conclusions}

While the observed enhancements of the leptonic 
and semileptonic $B$ meson decay rates involving a $\tau$ lepton relative to the expectations of the SM of electroweak interactions are intriguing,
their significance is not sufficient to unambiguously establish a violation of lepton universality at this time. However, the fact that these unexpected enhancements have been observed by three experiments operating in very different environments deserves further attention.

At present, the measurements are limited by the size of the available
data samples and uncertainties in the reconstruction efficiencies and background estimates.  It is not inconceivable that the experiments have underestimated these uncertainties, or missed a more conventional explanation.  Furthermore, while it is unlikely, it cannot be totally excluded that the theoretical SM predictions are not as firm as presently assumed. 
Currently, the experimenters are continuing their analysis efforts, refining their methods, enhancing the signal samples by adding additional decay modes, improving the efficiency and selectivity of the tagging algorithms, as well as the Monte Carlo simulations, and scrutinizing all other aspects of the signal extraction. 

In the near future, LHCb will make several important contributions, among them their first measurement of the $\BDtaunu$ decay, which will also improve results for $\BDstaunu$. Furthermore, the $\tau^{-} \to \pi^{-} \pi^{+} \pi^{-} \nu_{\tau}$ decay mode will be included.   
In addition, searches for lepton universality violation in 
semileptonic decays of other $B$ mesons and baryons are being planned.  
Beyond that, LHCb will continue to record data at the highest $pp$ collision energy available. By the end of 2017, the accumulated data sample is expected to increase by a factor of three.
In the longer term future, LHCb is planning to further enhance the data rate capability and record much larger event samples. 

At KEK in Japan, the $\epem$ collider is undergoing a major upgrade 
and is expected to enlarge the data sample by almost two orders of magnitude
over a period of about ten years. 
In parallel, the capabilities of the Belle detector are also being upgraded.
The operation of this new and more powerful detector is expected to start in 2018. 
The much larger event samples and the constrained $\BB$ kinematics will allow more precise measurements of kinematic distributions and detailed studies, for instance, of the $\tau$ polarization in
$\B \rightarrow D^{*} \tau \nu_{\tau}$ decays.
The feasibility of such a measurement was recently presented~
\cite{Hirose:2016wfn}.
For $\Bm \to \taum \nutb$ decays, which currently have statistical and systematic uncertainties of 30\% or more for individual measurements, the substantially larger data samples are expected to lead to major reductions in these uncertainties allowing more accurate assessments
of the compatibility with the SM predictions.
Detailed studies of the overall physics goals and precision measurements that can be achieved by Belle II and LHCb are ongoing.

In recent years, several experiments have examined decay rates and angular distributions for $\Bp$ decays involving a $K^{(*)+}$ meson and a lepton pair, 
$\Bp \to K^{(*)+}\mu^+ \mu^-$ and  $\Bp \to K^{(*)+} e^+ e^-$.   
In the framework of the SM these decays are very rare, since they involve 
$b \to s$ quark transitions.  LHCb~\cite{Aaij:2014ora} recently published a measurement of the ratio,
\begin{equation}
\label{eq:RK}
{\cal R}_K = \frac {{\cal B}(\Bp \to K^+ \mu^+ \mu^-)}
                        {{\cal B}(\Bp \to K^+ e^+  e^- )} 
                = 0.745 ^{+0.090}_{-0.074} \pm 0.036 , \
\end{equation}
a value that is 2.6 standard deviations below the SM expectation of about 1.0.
Earlier measurements by Belle~\cite{Wei:2009zv}, CDF~\cite{Aaltonen:2011ja}, and BABAR~\cite{Lees:2012tva} had significantly larger uncertainties and were fully consistent with lepton universality. Some theoretical models include new types of interactions that can explain this result. For instance, leptoquarks which can mediate this decay and result in higher rates for electrons than muons~\cite{Hiller:2014yaa,Becirevic:2016yqi}.
BABAR~\cite{Lees:2012tva}, LHCb~\cite{Aaij:2015oid} and 
Belle~\cite{Wehle:2016yoi} have analyzed angular distributions for the four decay modes and observed general agreement with SM predictions, except for local deviations, the most significant by LHCb at the level of 3.4 standard deviations.  Also here, more data are needed to enhance the significance of these measurements and find possible links to $\B$ decays involving $\tau$ leptons.
 
If the currently observed excess in the ratios \RD\ and \RDs\ is confirmed, experimenters will use their large data samples to measure properties of signal events and learn about the nature of the new particles and interactions that contribute to these decays~\cite{Sakaki:2014sea,Alonso:2016gym}.

In conclusion, we can expect much larger event samples from the upgraded LHCb and Belle experiments in the not too distant future.  
These data will be critical to the effort to 
understand whether the tantalizing results obtained to date are an early indication of beyond-the-SM physics processes or the result of larger-than-expected statistical or systematic deviations.
A confirmation of new physics contributions in these decays would shake the foundations of our understanding of matter and trigger an intense program of experimental and theoretical research.

\subsection*{Acknowledgements}
\label{sec:acknowledgements}
We recognize the contributions and dedication of our colleagues in the large international collaborations
supporting the operation of the BaBar (M.F.S., R.K., V.L.), Belle (T.K., Y.S.) and LHCb (G.C., B.H.)
detectors, the data processing and the data analyses on which the results presented in this Review are
based. None of this would have been achieved without the efforts of the teams at SLAC, KEK and CERN who
achieved excellent beam conditions and delivered high luminosities of the \epem and pp storage rings over many
years. We acknowledge support from the Organisation for Scientific Research (NWO) of the Netherlands, the US
National Science Foundation and Department of Energy, the Natural Sciences and Engineering Research Council
(NSERC) of Canada, the Excellence Cluster of the DFG of Germany: Origin and Structure of the Universe, and the
Japan Society for the Promotion of Science (JSPS).



\begin{thebibliography}{10}
\bibitem{charge}
Unless stated otherwise, the inclusion of charged-conjugate states and decay
  modes is implied here and in the following.
\bibitem{helicity}
Helicity refers to the component of the angular momentum of a particle parallel
  to the direction to its momentum.
\end{thebibliography}
\end{document}